\documentclass[twocolumn,showpacs,preprintnumbers,amsmath,amssymb,nofootinbib]{revtex4-1}

\usepackage{graphicx}
\usepackage{dcolumn}
\usepackage{bm}
\usepackage{amsmath,dsfont}
\usepackage{color}

\def\IR{{\mathds{R}}}

\def\H{{\mathcal{E}}}
\def\I{{\mathcal{I}}}
\def\H{{\mathcal{H}}}

\def\Q{{\mathcal{Q}}}

\def\p{{\partial}}

\def\hx{{\hat{r}}}
\def\vn{{\hat{n}}}

\def\vJ{{\vec{J}}}
\def\vL{{\vec{L}}}

\def\vK{{\vec{K}}}

\def\vA{{\vec{A}}}

\def\vD{{\vec{D}}}

\def\vC{{\vec{C}}}

\def\vpi{{\vec{\pi}}}

\def\smallover#1/#2{\hbox{$\textstyle\frac{#1}{#2}$}}
\def\IR{{\mathds{R}}}

\def\M{{\mathcal{M}}}

\def\b{{\beta}}

\def\vA{{\vec{A}}}
\def\vK{{\vec{K}}}
\def\vJ{{\vec{J}}}
\def\vPi{{\vec{\Pi}}}
\def\g{{\mathcal{g}}}

\def\b{{\beta}}
\def\d{{\delta}}

\def\vx{{\vec{x}}}

\def\vn{{\vec{n}}}
\def\va{{\vec{a}}}

\def\vnabla{{\vec{\nabla}}}
\def\hx{{\frac{\vec{x}}{r}}}

\def\beq{\begin{equation}}
\def\eeq{\end{equation}}
\def\beqa{\begin{eqnarray}}
\def\eeqa{\end{eqnarray}}
\def\g{\gamma}

\def\nn{\nonumber}

\def\IR{{\mathds{R}}}

\def\E{{\mathcal{E}}}
\def\I{{\mathcal{I}}}
\def\H{{\mathcal{H}}}

\def\Q{{\mathcal{Q}}}

\def\p{{\partial}}

\def\hx{{\hat{r}}}

\def\vn{{\hat{n}}}

\def\vJ{{\vec{J}}}
\def\vL{{\vec{L}}}

\def\vK{{\vec{K}}}

\def\vA{{\vec{A}}}

\def\vD{{\vec{D}}}

\def\vC{{\vec{C}}}

\def\vpi{{\vec{\pi}}}

\def\smallover#1/#2{\hbox{$\textstyle\frac{#1}{#2}$}}

\def\M{{\mathcal{M}}}

\def\S{{\mathcal{S}}}

\def\b{{\beta}}

\def\k{{\kappa}}
\def\vA{{\vec{A}}}
\def\vK{{\vec{K}}}
\def\vJ{{\vec{J}}}
\def\vPi{{\vec{\Pi}}}
\def\g{{\mathcal{g}}}

\def\L{{\mathcal{L}}}

\def\b{{\beta}}
\def\d{{\delta}}

\def\vx{{\vec{x}}}

\def\vn{{\vec{n}}}
\def\va{{\vec{a}}}

\def\vnabla{{\vec{\nabla}}}
\def\hx{{\frac{\vec{x}}{r}}}

\def\beq{\begin{equation}}
\def\eeq{\end{equation}}
\def\beqa{\begin{eqnarray}}
\def\eeqa{\end{eqnarray}}
\def\g{\gamma}

\def\nn{\nonumber}

\usepackage{color}

\begin{document}

\preprint{
}

\title{Dynamical symmetries of generalized Taub-NUT and multi-center metrics}

\author{\large J.-P.~Ngome
}
\affiliation{
International Basic Sciences Programme (IBSP), UNESCO,
1 rue Miollis, 75015 Paris, France.
}
\email{jj.ngome-abiaga-at-unesco.org
}

\begin{abstract}
\noindent
Hidden symmetries of generalized Kaluza-Klein-type metrics are studied using van Holten's systematic analysis \cite{vH} based on Killing tensors. Applied to generalized Taub-NUT metrics, Kepler-type symmetries with associated Runge-Lenz-type conserved quantities are constructed. In the multicenter case, the subclass of two-center metrics gives rise to a conserved Runge-Lenz-type scalar, while no Kepler-type constant of the motion does exist for non aligned $(N\geq3)$-centers. We also investigated the diatomic molecule system of Wilczek et al. where ``truly'' non-Abelian gauge fields mimicking monopole-like fields arised. From the latter system we deduced a new conserved charge.
\end{abstract}


\maketitle


\section{Introduction to Kaluza-Klein Theories}
Kaluza-Klein (KK) theories have been extensively studied as schemes attempting to unify gravitation and gauge theory \cite{Kaluza,Klein}, through the physical assumption that the world admits, in addition to 4D space-time, an unobservable extra dimension. Thus, ordinary general relativity in five dimensions is considered to possess a local $\,U(1)\,$ gauge symmetry arising from a ``vertical'' translation along the hidden extra dimension. 

A perfect illustration of the KK framework has been given by the Sorkin, then Gross and Perry solution of the vacuum Einstein equation, involving an Abelian monopole potential \cite{Sork,GrossPerry} which carries, unexpectedly, Kepler-type dynamical symmetries. Further examples  are provided by the multi-center metrics \cite{GRub,FH,GR,GW} for which similar hidden symmetry properties have been revealed. 

Let us consider the 5D KK metric tensor, 
\beq
\begin{array}{cc}\displaystyle 
g_{AB} = \left(\begin{array}{cc} \g_{\mu\nu}+VA_{\mu}A_{\nu}  & \quad A_{\mu}V\\[4pt] A_{\nu}V  & V\end{array}\right),\\[12pt]
A, B=0,\cdots,4;\qquad\mu,\,\nu=0,\cdots,3;\qquad\displaystyle A_{0}=0\,,
\end{array}\label{KK5metric}
\eeq
where the 5D manifold can be viewed as a direct product of a 4D space-time (where $x^0$ is the time coordinate) with an unobservable space-like loop, $\M^{4}\otimes \S^{1}$, and where all components of $g_{AB}$ are independent of the extra coordinate $x^4$. Here $\g_{\mu\nu}$ is the metric of the 4D manifold $\M^4$. From (\ref{KK5metric}), the dynamics of a classical point-like test particle of unit mass is given by the 5D geodesic motion,
\beq
\frac{d^{2}x^{A}}{d\tau^{2}}+\Gamma^{A}_{BC}\frac{dx^{B}}{d\tau}\frac{dx^{C}}{d\tau}=0\,,
\label{GeoEq}
\eeq
where $\tau$ denotes the proper time. Using the effective theory (\ref{KK5metric}) in (\ref{GeoEq}), a routine calculation yields the equations of the motion,
\beq
\left\lbrace\begin{array}{ll}\displaystyle
\frac{d}{d\tau}\big(VA_{\mu}\frac{dx^{\mu}}{d\tau}+V\frac{dx^4}{d\tau}\big)=\frac{dq}{d\tau}=0\,,
\\[12pt]
\displaystyle
\frac{d^{2}x^{\mu}}{d\tau^2}+\Gamma^{\mu}_{\nu\lambda}\frac{dx^{\nu}}{d\tau}\frac{dx^{\lambda}}{d\tau}-qF^{\mu}_{\lambda}\frac{dx^{\lambda}}{d\tau}-\frac{q^2}{2}\frac{\p^{\mu}V}{V^2}=0\,.
\end{array}
\label{EqGeoM}\right.
\eeq
Here the $x^\mu$ are the coordinates on the 4D Manifold $\M^4$ while $x^4$ represents the ``extra" coordinate. The first equation in (\ref{EqGeoM}) tells us that the ``charge'',
\beq
q=V\left(A_{\mu}\frac{dx^{\mu}}{d\tau}+\frac{dx^4}{d\tau}\right)\,,
\eeq
is conserved along the 5D geodesics. The latter can also be viewed as being associated with translations in the ``extra'' direction, generated by the Killing vector $\p_{x^4}$.  The second equation in (\ref{EqGeoM}) is a 4D geodesic equation involving an interaction with the scalar field $V$  in addition to the Lorentz force . See \cite{Kerner2, Kibble,Trautman2} for further references. 
\newline

The non-Abelian generalization of the 5D KK approach was given by Kerner \cite{Kerner} through generalizing the previous 5D manifold to a (4+d)-dimensional one, namely \break$\,
\M=\M^{4}\otimes \S^{d}\,$ whose base space $\M^{4}$ is the usual space-time and where $\S^{d}$ represents an unobservable $d$-dimensional extra space.  Here the (4+d)D diffeomorphism symmetry is broken to 4D infinitesimal coordinate transformations augmented with translations along the extra dimensions,
\beq
x^{\mu}\rightarrow x^{\mu}+\d x^{\mu}\,,\quad y^{a}\rightarrow y^{a}+f^{i}(x^{\nu})\xi_{i}^{a}(y)\,,\label{KKNAtransf}
\eeq
where $f^{i}(x^{\nu})$ are infinitesimal functions and $\xi_i^{a}$ denoting the isometry generators on the compact manifold $\S^d$. The (4+d)D generalized metric, invariant under (\ref{KKNAtransf}), then reads
\beq
\tilde{g}_{CD} =  \left(\begin{array}{cc} \g_{\mu\nu}+\k_{ab}B_{\mu}^aB_{\nu}^{b}  & \quad B_{\mu}^{b}\k_{ba}\\[4pt] \k_{ab}B_{\nu}^{a}  & \k_{ab}\end{array}\right)\,,\nn
\eeq 
where $\k_{ab}$ represents the $\,SU(d-1)\,$ invariant metric on $\S^{d}$, and $\,
\,B_{\mu}^{a}=A_{\mu}^{b}\,\xi^{a}_{b}\,
\,$
includes the $SU(d-1)$ Lie algebra valued one-form $\,A_{\mu}^{b}\,$, identified as a \emph{Yang-Mills field}.  

We can now deduce the geodesic equations which yield the equations of motion of an isospin-carrying particle in a curved space plus a Yang-Mills field,
\beq
\left\lbrace\begin{array}{ll}\displaystyle
\mathcal{D}_{\tau}\I_c=\frac{d\I_{c}}{d\tau}-\I_{a}\varepsilon_{\;bc}^{a}A^{b}_{\mu}\frac{dx^{\mu}}{d\tau}=0\,,\\[8pt]
\displaystyle
\frac{d^{2}x^{\b}}{d\tau^{2}}+\Gamma^{\b}_{\mu\nu}\frac{dx^{\mu}}{d\tau}\frac{dx^{\nu}}{d\tau}+\g^{\nu\b}F_{\mu\nu}^{b}\I_{b}\frac{dx^{\mu}}{d\tau}=0\,.
\end{array}\right.
\label{KKNAKerner-Wong}
\eeq
Analogously to the Abelian case (\ref{EqGeoM}), the first equation in the pair (\ref{KKNAKerner-Wong}) identifies the \emph{classical isospin variable}, $$\,\displaystyle
\I_a=\k_{ab}\left({dy^{b}}/{d\tau}+A^{b}_{\nu}{dx^{\nu}}/{d\tau}\right)\,,
\,$$ which is parallel transported and describes the motion in non-Abelian internal space. The  isospin is analogous to the electric charge in the Abelian KK theory. Indeed, it is also obtained by  contracting the Killing vector which generates the ``vertical'' translations with the direction field of the geodesic. The second equation in (\ref{KKNAKerner-Wong}) still describes the motion in 4D real space. Note here the generalized Lorentz force 
\beq
\g^{\nu\b}F_{\mu\nu}^{b}\I_{b}({dx^{\mu}}/{d\tau})\nn
\eeq
due to the Yang-Mills field and with the electric charge  replaced by the isospin. 

The equations (\ref{KKNAKerner-Wong}) are known as the \emph{Kerner-Wong equations}, as they were also obtained by Wong \cite{Wong} by ``dequantizing'' the Dirac equation. They can be derived from a variational principle  \cite{Bal}; alternatively, they can be
studied using a symplectic approach \cite{DHInt,Feher:1986*}.

\section{van Holten's method to derive the constants of the motion}

Now, inquiring about the symmetries of the KK-type metrics, we recall that constants of the motion denoted as $\,Q\,$, which are polynomial in the momenta, can be derived following van Holten's algorithm \cite{vH,Ngome1}. The recipe  is to expand $\,Q\,$ into a power series of the covariant momentum,
\beq\displaystyle
Q= C+C^i\,\Pi_i+\frac{1}{2!}\,C^{ij}\,\Pi_i\Pi_j+\frac{1}{3!}\,C^{ijl}\,\Pi_i\Pi_j\Pi_l+\cdots\,\nn
\eeq
and to require $Q$ to Poisson-commute with the Hamiltonian augmented with an effective potential,
$\,\left\lbrace\Q\,,\mathcal{H}={\vec{\Pi}^{2}}/{2}+G(\vx)\right\rbrace=0$. This yields the series of constraints,
\beqa
\left\lbrace
\begin{array}{lllll} 
C^m\partial_mG=0& \hbox{o(0)} & \\[7pt]
\partial_n C=qF_{nm}C^m+C_n^{\;m}\p_mG & \hbox{o(1)} &\\[7pt]
\mathcal{D}_{\left(i\right.}C_{\left.l\right)}=q(F_{im}C_l^{\,m}+F_{lm}C_i^{\,m})+C_{il}^{\;\;k}\p_kG & \hbox{o(2)} &\\[7pt]
\mathcal{D}_{\left(i\right.}C_{\left. lj\right)}=qF_{im}C_{lj}^{\;\,m}+qF_{jm}C_{il}^{\;\,m}\\[4pt]
\qquad\qquad+qF_{lm}C_{ij}^{\;\,m}+C_{ijl}^{\;\;\;m}\p_mG & \hbox{o(3)} &
\\
\cdots\cdots
\end{array}\right.
\label{ConsTraints}
\eeqa
Here the zeroth-order constraint can be interpreted as a consistency condition for the effective potential. It is worth noting that the expansion can be truncated at a finite order provided some higher-order constraint reduces to a Killing equation,  
\beq
\mathcal{D}_{\left(i_1\right.}C_{\left. i_2\;\cdots\;i_n \right)}=0\,,\label{KillingNAbis}
\eeq
where the covariant derivative is constructed with the Levi-Civita connection so that
\beq
\displaystyle{\mathcal{D}_iC^j =\partial_iC^j +\Gamma^j_{\;ik}\,C^k}\,.
\eeq
Then $\,\displaystyle{C_{i_1\cdots i_p}=0}\,$ for all $\,p\,\geqslant\,n\,$ and the constant of the motion takes the polynomial form,
\beq
Q=\sum_{k=0}^{p-1}\,\frac{1}{k!}\,C^{i_1\cdots i_k}\,\Pi_{i_1}\cdots\Pi_{i_k}\,.\label{ConstExp}
\eeq\\
It is worth noting that apart from  zeroth-order conserved charges which are independent of the  covariant momentum, all order-\textit{n} invariants are deduced by the van Holten  method (\ref{ConsTraints}) involving rank-\textit{n} Killing tensors of the curved manifold \cite{Killing}. Generating symmetries using Killing tensors  was first advocated by Carter in the context of the Kerr metric \cite{Carter}. 

In our set of constraints (\ref{ConsTraints}), a given Killing tensor is the highest-order coefficient of the expansion (\ref{ConstExp}), allowing us to solve the truncated series of constraints and thus generating a conserved quantity. The intermediate-order constraints in (\ref{ConsTraints}) then determine the other coefficient-terms of the invariant (\ref{ConstExp}).

In what follows, our strategy    
 will be to find conditions for lifting to the ``Kaluza-Klein'' $4$-space those Killing tensors which generate, in flat space,
  the conserved angular momentum and the Runge-Lenz vector of planetary motion, respectively.

\section{Hidden symmetries of generalized Taub-Nut metrics}

Let us first  investigate the symmetries of the [Abelian] Kaluza-Klein monopole. The latter, obtained by imbedding the Taub-NUT gravitational instanton into Kaluza-Klein theory \cite{Sork, GrossPerry,GM},  
provides us with an exact solution  of  four-dimensional Euclidean gravity approaching the vacuum solution at spatial infinity,
\beq
\left\lbrace\begin{array}{ll}\displaystyle 
ds^2=g_{ij}(\vx)dx^idx^j+h(\vx)\big(dx^4+A_kdx^k\big)^2\\[8pt]\displaystyle
g_{ij}(\vx)=f(\vx)\delta_{ij}(\vx)
\end{array}\right.
\label{metricTN}
\eeq
where $\,f(\vx)\,$, $\,h(\vx)\,$ and $\,A_k\,$ are real functions and the gauge potential of an Abelian magnetic field, respectively. The Lagrangian function associated reads 
\beq
\mathcal{L}=\frac{1}{2}\,f(\vx)\,\dot{\vx}^{\,2}+\frac{1}{2}f^{-1}(\vx)\,\big(\,\frac{dx^4}{dt}+A_k\,\frac{dx^k}{dt}\,\big)^2-U(r),\nn\eeq
where $U(r)$ is an external scalar potential.

Inspired by Kaluza's hypothesis, as the fourth dimension $x^4$ is considered to be cyclic, we use the conservation of the ``vertical'' component of the momentum interpreted as a conserved electric charge,
\beq\displaystyle
p_4=h(\vx)(dx^4/{dt}+A_k{d x^k}/{dt})=q\,,\nn
\eeq
to reduce the four-dimensional problem to one in three dimensions, where we have strong candidates for the way dynamical symmetries act \cite{Ngome:2009pa}. Then, the lifting problem can be conveniently solved using the Van Holten technique (\ref{ConsTraints}) based on Killing tensors. The geodesic motion on the 4-manifold therefore projects onto the curved 3-manifold with the metric $g_{ij}(\vx)$, augmented with an effective potential,
\beq
V(\vx)={q^{2}}/{2h(\vx)}+U(r)\,,\nn
\eeq
such as
\beq
\left\lbrace\begin{array}{ll}\displaystyle 
\dot{x}^i=g^{ij}\Pi_j\,,\quad\Pi_j=p_j-q\,A_j\,,
\\[8pt]\displaystyle
\dot{\Pi}_{i}=
qF_{ij}\dot{x}^j-\partial_i\displaystyle{V} +\Gamma^k_{ij}\Pi_k\dot{x}^j\,.
\end{array}\right.
\label{Lorentz}
\eeq
Note that the Lorentz equation in (\ref{Lorentz}) involves also in addition to the monopole and potential terms, a curvature term which is quadratic in the velocity.

Let us now focus Kaluza-Klein-type metrics (\ref{metricTN}) 
 whose hidden symmetries have been extensively investigated \cite{GRub,FH,GM,Ngome:2009pa,GR,CFH1,VVHH1, Japs1, Japs2,Valent,VVHH2,CoHo,Visi2,VamanVisi,Krivonos, Ballesteros,K-L,K-L-S,Nerses,Marquette2}. For geodesic motion on hyperbolic space, for instance, Gibbons and Warnick \cite{GW} found a large class of systems admitting such hidden symmetries. 

We assume that the metric (\ref{metricTN}) is radial,
\beq
f(\vx)=f(r), \quad h(\vx)=h(r)\,,\nn
\eeq
and the magnetic field is that of a Dirac monopole of charge $g$. In that event, the generator of spatial rotations,
\beq
C_i=g_{ij}(r)\,\epsilon^j_{\;kl}\;n^k\;x^l\,,
\eeq
applied to (\ref{ConsTraints}) provides us with the conserved angular momentum involving the typical monopole term,
\beq
\displaystyle
\vJ=\vx\times\vPi-qg\hx\,.
\eeq
Turning to second order Runge-Lenz-type conserved quantities, we use the rank-2 Killing tensor,
\beq\displaystyle
C_{ij}= 2g_{ij}(r)n_kx^k-g_{ik}(r)n_jx^k-g_{jk}(r)n_ix^k\,,
\label{RL1}
\eeq
inspired by its form in the Kepler problem. We therefore deduce from (\ref{ConsTraints}) that:
\begin{enumerate}
\item
 For the original Taub-NUT case \cite{Sork,GrossPerry}  with no external scalar potential,
\beq
f(r)=\frac{1}{h(r)}
=1+\frac{4m}{r},
\label{TNcase}
\eeq
where $m$ is real\footnote{Monopole scattering corresponds to $m=-1/2$ \cite{GM}.} \cite{GR,FH} we obtain, for the energy $\E$ and the charge $\,g=\pm 4m\,$, the conserved Runge-Lenz vector,
\beq
\vec{K}=\vPi\times\vec{J}-4m\left(\mathcal{E}-q^{2}\right)\hx\,.\label{KKRL}
\eeq
\item  Lee and Lee \cite{LeeLee} argued that for monopole
scattering with independent Higgs expectation values, the geodesic
Lagrangian derived from (\ref{metricTN}) should be replaced by $\L\to\L-W(r)$,
where
\beq
W(r)=\frac{1}{2}\,\frac{a^{\;\,2}_{0}}{1+\displaystyle\frac{4m}{r}}\ .
\label{LLpot}
\eeq
It is easy to see that this addition merely shifts the
value of the energy by a constant $a^{\;\,2}_{0}/2$, so the previously
found Runge-Lenz vector (\ref{KKRL}) is still valid.
\item
The metric associated with winding strings \cite{GR2},
\beq
f(r)=1,
\qquad 
h(r)=\frac{1}{\big(1-\displaystyle\frac{1}{r}\big)^2}\,,
\eeq
with charge $\,g=\pm 1\,$, leads to the conserved Runge-Lenz vector,
\beq\displaystyle{
\vec{K}={\vPi}\times\vec{J}-q^{2}\,\hx}\,.
\eeq
\item The extended Taub-NUT metric \cite{Japs1, Japs2}
\begin{eqnarray}\begin{array}{cc}\displaystyle{
f(r)= b+\frac{a}{r},
\quad 
h(r)= \frac{a\,r+b\,r^2}{1+d\,r+c\,r^2}}\,,
\end{array}
\label{japexp}
\end{eqnarray}
with $(a,\, b,\, c,\, d)$ real.
With no external scalar potential and charge $\,g=\pm 1\,$, we have the conserved Runge-Lenz vector,
\beq
\vec{K}=\vPi\times\vec{J}-(a\,
\mathcal{E}-\frac{1}{2}\,d\,q^{2})\hx\,.
\eeq
\item Considering the oscillator-type metric discussed by Iwai and Katayama \cite{Japs1}, the functions $\,f(r)\,$  and $\,h(r)\,$ take the particular form
\beq
\displaystyle
f(r)= b+ar^2,\;\, h(r)= \frac{ar^4+br^2}{1+cr^2+dr^4}\,.
\eeq
A direct calculation leads to the following Runge-Lenz-type vector \cite{Japs1} ,
\beq
\vK=\left(b+ar^2\right)\,\dot{\vx}\times\vJ+\beta\,\frac{\vx}{r}\,,
\eeq
conserved only for a scalar potential of the form
\beq
\displaystyle
W(r)=(\frac{q^2g^2}{2r^2}+\frac{\beta}{r})\frac{1}{f(r)}-\frac{q^2}{h(r)}\,.
\eeq
\end{enumerate}

\section{N-center metrics with scalar constants of the motion}

The multi-center metrics in which we are interested here are Euclidean vacuum solutions of the Einstein equations with self-dual curvature; they can  be viewed as generalizations of the previously investigated Taub-NUT metrics.

 Let us consider a particle moving in  Gibbons-Hawking space \cite{GH}. The Lagrangian function associated with this dynamical system, derived from (\ref{metricTN}),  is 
\begin{eqnarray*}\displaystyle{
\mathcal{L}=\frac{1}{2}\,f(\vx)\,\dot{\vx}^{\,2}+\frac{1}{2}f^{-1}(\vx)\,\big(\,\frac{dx^4}{dt}+A_k\,\frac{dx^k}{dt}\,\big)^2-U(\vx)},
\end{eqnarray*}
where the functions $\,f(\vx)\,$ obey the ``self-dual'' condition
$\,\displaystyle{
\vnabla\,f(\vx)=\pm\vnabla\times\vA}\,$. 
$f(\vx)$ satisfies therefore the three-dimensional Laplace equation,
\begin{eqnarray*}\displaystyle{
\Delta\,f(\vx)=0}\,,
\end{eqnarray*}
whose most general solution is given by
\beq\displaystyle
f(\vx)=f_{0}+\sum_{i=1}^{N}\frac{m_{i}}{\vert\vx-\va_{i}\vert},\;(f_{0},\,m_{i})\in \IR^{N+1}.\nn
\eeq
The multicenter metric admits multi-NUT singularities so that the position of the \textit{ith} NUT singularity with the charge $\,m_{i}\,$ is $\,\va_{i}\,$. These singularities can be removed provided  to have all  NUT charges equal. In this case, the cyclic variable $\,x^{4}\,$ is periodic with  range $\,\displaystyle{0 \leq x^{4}\leq 4\pi m}\,$. We are thus interested in the projection of the motion  on the curved 3-manifold whose metric is,
\begin{eqnarray}\displaystyle{
g_{jk}(\vx)=(
f_{0}+\sum_{i=1}^{N}\frac{m_{i}}{\vert\vx-\va_{i}\vert})\,\delta_{jk}}\,.\label{MCmetric}
\end{eqnarray}
For simplicity, we limit ourselves to two-center metrics, 
\beq\displaystyle{
f(\vx)=f_{0}+\frac{m_{1}}{\vert\vx-\va\vert}+\frac{m_{2}}{\vert\vx+\va\vert}}\,.
\eeq
Turning to rotational symmetry, the rank-1 Killing tensor satisfying the second-order equation in (\ref{ConsTraints}),
\beq\displaystyle
C_i=g_{im}\epsilon^m_{\;\;\;l k}{\hat{a}^l}x^k\,, \quad\hat{a}=\va/a\,,\label{Killing1MC}
\eeq
generates rotational symmetry around the axis through the two centers. The corresponding conserved quantity,
\beq\displaystyle
\mathcal{J}_a=(\vx\times\vPi)\cdot{\hat{a}}-q(f(\vx)-f_0)\vx\cdot\hat{a}+\frac{qam_1}{\vert \vx-\va\vert}-\frac{qam_2}{\vert \vx+\va\vert}\,,\nn
\eeq
is  the \emph{projection of the angular momentum} onto the axis of the two centers.

Now we study quadratic conserved quantities by considering the reducible rank-2 Killing tensor,
\beq\begin{array}{ll}\displaystyle
C_{ij}={2}g_{im}g_{jn}\epsilon^m_{\;\,l k}\epsilon^n_{\;p q}\hat{a}^l\hat{a}^px^kx^q+{2}g_{il}g_{jm}\hat{a}^l\hat{a}^m\,,\label{KillTensor}
\end{array}\eeq
which is a symmetrized product of the Killing-Yano tensors, $\displaystyle{C_{i}=g_{im}\epsilon^m_{\;\;\;l k}\;\hat{a}^l\,x^k}$ generating rotations around the axis of the two centers and $\,\displaystyle{{C}_j=g_{jm}\hat{a}^m }\,$ generating spatial translation along the axis of the two centers. Injecting (\ref{KillTensor}) into the system of constraint (\ref{ConsTraints}) yields, for vanishing effective potential $U=0$, the Casimir, which combines the \emph{square of the projected angular momentum with the
square of the component of the covariant momentum along the axis},
\begin{eqnarray}\displaystyle{
Q=\mathcal{J}_a^{2}+\Pi^2_a}\,.
\end{eqnarray}
As expected from the construction of the associated Killing tensor \cite{GRub}, the obtained conserved quantity is a combination of two constant of the motion \cite{Valent}.

Now introducing into (\ref{ConsTraints}) the rank two Killing tensor generating Kepler-type dynamical symmetry provides us with the conserved \emph{scalar},
\beq\displaystyle
K_{a}=\left(\vPi\times\vJ\right)\cdot\frac{\va}{a}+\frac{\beta}{q}\,\left(\mathcal{L}_{a}-\mathcal{J}_{a}\right)
\,,\label{RLScalar}
\eeq
where $\,\beta\in\IR$. $\,K_a\,$ therefore represents, for the two-center metrics (\ref{MCmetric}), a conserved Runge-Lenz-type \emph{scalar} for particle  motion confined onto the ``Appolonius" two-sphere\footnote{This sphere was known already by Apollonius of Perga in the 2th century BC.} \cite{Ngome:2009pa} of center  at $\,\va\,\rho\,$ and with radius $\,\displaystyle{R = a\,\sqrt{\rho^{2}-1}}\,$, provided the effective potential is of the form,
\beq
W=
\frac{q^{2}}{2}(f(\vx)-f_0)^{2}
+\beta(f(\vx)-f_0)+\gamma\,
\label{FormPot3}
\eeq
$\gamma$ a constant, which satisfies the consistency condition given by the zeroth-order constraint of (\ref{ConsTraints}). 

It is worth mentioning that a scalar Runge-Lenz-type conserved quantity does exist only for a particle moving along the axis of the two centers, or for motions confined onto the ``Apollonius" two-sphere defined above. In the Eguchi-Hanson case $\,m_{1}=m_{2}\,$, and the $2$-sphere is replaced by the median plane of the two centers  \cite{Ngome:2009pa}.
\newline
\noindent

The $(N\geq3)$-centers metrics of the form
\begin{eqnarray}\displaystyle{
g_{jk}(\vx)=f(\vx)\delta_{jk}=(
f_{0}+\sum_{i=1}^{N\geq3}\frac{m_{i}}{\vert\vx-\va_{i}\vert})\,\delta_{jk}}\,
\label{3Cmetric}
\end{eqnarray}
 can also be investigated, but they carry no Runge-Lenz-type symmetry for $N\geq3$ non aligned centers. To obtain this result, let us first generalize the rank-2 Killing tensor (\ref{RL1}) generating the Runge-Lenz vector as 
\beq
C_{ij}=2g_{ij}(\vx)n_kx^k-g_{ik}(\vx)n_jx^k-g_{jk}(\vx)n_ix^k\,.
\label{KillT2}
\eeq
Requiring the Killing equation to be satisfied as
\beq\displaystyle
\mathcal{D}_{\left( k\right.}C_{\left. ij\right)}=0\,,
\eeq
a tedious calculation then provides us with the  condition
\beq
\vn\times\left(\vx\times\vnabla\,f(\vx)\right)=0\,.
\eeq
For the metric (\ref{3Cmetric}) above this requires
\beq
\displaystyle\sum_{i=1}^{N}\frac{\big(\vn\cdot\vx\big)\va_{i}-\big(\vn\cdot\va_{i}\big)\vx}{\vert\vx-\va_{i}\vert^{3}}=0\,,\label{CCC}
\eeq
which can not be satisfied for more than two non-aligned centers. The only possibility we get for (\ref{CCC}) to be satisfied by more than two centers is to have all of them in the same alignement describing a straight line of centers.

\section{The non-Abelian case~: the diatomic molecule}

 In Ref. \cite{MSW} Moody, Shapere and Wilczek have shown that nuclear
motion in a diatomic molecule can be described,  in the Born-Oppenheimer approximation, by an effective non-Abelian gauge field of ``hedgehog'' form,
\beq
{A}_i^{\;a}=(1-\kappa)\epsilon_{iaj}\,
\frac{x^j}{r^2}\,,
\quad 
{F}^{\;a}_{ij}=(1-\kappa^2)\epsilon_{ijk}
\frac{x^kx^a}{r^4}\,.
\label{diatfields}
\eeq
This gauge field mimics the structure of that of a non-Abelian monopole \cite{tHooft,Polyakov}. Note here \textit{the unquantized constant real factor $\,(1-\kappa^2)\,$ }.
The potential  (\ref{diatfields}) becomes that of a Wu-Yang [i.e., an imbedded Dirac] monopole of unit charge when $\k=0$; for other values of $\kappa$, it is a truly non-Abelian configuration -Ñ except for $\k=\pm1$, when it is a gauge transform of the vacuum.

Now we investigate the symmetries of an isospin-carrying particle carrying unit charge, evolving in the monopole-like field of the diatom (\ref{diatfields}) plus a scalar potential. The Hamiltonian describing the dynamics of this particle is finally expressed as
\beq
\H=\frac{\vpi^{2}}{2}+V(\vx,\,\vpi,\,\I^{a})\,,\quad\pi_{i}=p_{i}-{A}_{i}^{a}\,\I^{a}\,.\nn
\eeq 
Defining the covariant Poisson-brackets as 
\beq
\begin{array}{ll}\displaystyle
\big\{M,N\big\}=D_jM\frac{\p N}{\p \pi_j}-\frac{\p M}{\p \pi_j}D_jN
+\I^a{F}^{\;a}_{jk}\frac{\p M}{\p \pi_j}\frac{\p N}{\p \pi_k}\\[8pt]\displaystyle
\qquad\qquad\quad-\epsilon_{abc}\frac{\p M}{\p \I^a}\frac{\p N}{\p \I^b}\I^c\,,\nn
\end{array}
\eeq
the non-vanishing brackets are
\beq
\{x^{i},\pi_{j}\}=\d^{i}_{j}\,,\quad\{\pi_{i},\pi_{j}\}=\I^a{F}_{ij}^{a}\,,\quad\{\I^{a},\I^{b}\}=-\epsilon_{abc}\I^{c}\,.\nn
\eeq
The equations of motion governing an isospin-carrying particle in the static non-Abelian gauge field (\ref{diatfields}) read
\beq\left\lbrace\begin{array}{ll}\displaystyle
\ddot{x}_{i}-\I^a{F}_{ij}^{a}\,\dot{x}^{j}+D_{i}V=0\,,
\\[6pt]\displaystyle
\dot{\I}^{a}+\epsilon_{abc}\,\I^{b}({A}_{j}^{c}\,\dot{x}^{j}-\frac{\p V}{\p \I^{c}})=0\,.
\end{array}\right.\label{DiatEqM}
\eeq
These equations generalize the Kerner-Wong equations
(\ref{KKNAKerner-Wong}) to an additional scalar potential.

Turning to the conserved quantities constructed with the van Holten algorithm, the zeroth-order conserved charge which used-to-be interpreted as electric charge for $\kappa\neq0$, 
\beq
\displaystyle Q=\frac{\vx\cdot\vec{\I}}{r}\,,
\eeq
is \emph{not more covariantly conserved} in general,
\beq
\big\{Q,\H\big\}=\vpi\cdot \vD Q, 
\quad
D_jQ=\frac{\kappa}{r}(\I^j-Q\frac{x_j}{r}).
\label{noecharge}
\eeq
Detailed calculation shows that the equation $\,D_jQ=0\,$ can only be solved, for an imbedded Abelian monopole field, when $\,\kappa=0,\pm1\,$.
 
Nor is $Q^2$  conserved,
$
\big\{Q^2,\H\big\}=2\kappa Q(\vpi\cdot\vD Q)\,.
$
but note that,
unlike $Q^2$, the length of the isospin, $\I^2$, \emph{is} conserved, 
\beq\{\H,\I^2\}=0\,.\eeq
These results are consistent with those in \cite{HR85}.

Turning to linear conserved quantities, we use the Killing vector generating spatial rotations,
\beq
\vC=\vn\times\vx\,,
\eeq
to build up the conserved angular momentum,
\beq\displaystyle
\vJ=\vx\times\vpi-(1-\kappa)\,Q\,\hx-\kappa\vec{\I}.\,\label{diatangmomJ}
\eeq

Moody, Shapere and Wilczek \cite{MSW} did find this  expression for $\kappa=0$ but, as they say it, ``they are not aware of a canonical derivation
when $\kappa\neq0$''.  Our construction here is an alternative to that of Jackiw \cite{Jdiat}, who obtained 
it using the method of Ref. \cite{JackiwManton}.

We now consider the rank-$\,2\,$ Killing tensor,
\beq
C_{ij}=2\d_{ij}x^2-2x_{i}x_{j}\,,\label{KillingDiatom2}
\eeq
which satisfies the third-order constraint in (\ref{ConsTraints}). The Killing tensor (\ref{KillingDiatom2}) thus yields the \emph{conserved} Casimir,
\beq
L^2=\big(\vx\times\vpi\big)^{2}=x^2\vpi^2-\big(\vx\cdot\vpi\big)^2\,,\label{Casimir1}
\eeq
which is the \emph{square} of  \emph{non-conserved orbital angular momentum}, $\vL=\vx\times\vpi\,$. Since $\,J^2\,$ and $\,L^2\,$ are both conserved,  the new charge,
\beq
\Gamma=J^2-L^2=\big(1-\k\big)^2Q^2-\k^2\vec{\I}^2-2\k\vJ\cdot\vec{\I}\,,\label{Gamma}
\eeq
is conserved for motion in the monopole-like field of diatomic molecule. \textit{It is worth noting that the charge $\,\Gamma\,$ becomes, in the Abelian limit  $\,\k=0\,$,  the square of conserved electric charge}. 

Just like $\,\vJ\,$, $\,J^2\,$ and $\,L^2\,$, the charge $\,\Gamma\,$ is conserved for any radially symmetric potential, $\,V(r)\,$. $\,\Gamma\,$ can also be obtained by using the Killing vector,
\beq
\vC=2\k\big(\vx\times\vec{\I}\big)\,,
\eeq
in the van Holten algorithm (\ref{ConsTraints}). We note at last, that no Runge-Lenz-type conserved
quantity could be found in this case except in the Abelian case, cf. \cite{Ngome1}.
\section{Conclusion}

We studied the  geodesic motion of a particle in Kaluza-Klein-type monopole spaces and in its Gibbons-Hawking  generalization. As illustrations, we  treated in detail the generalized Taub-NUT metrics, for which we derived Runge-Lenz-type vectors. We considered the subclass of two-center metrics into which a conserved Runge-Lenz-type scalar has been revealed in the special case of motions confined onto a particular ``Apollonius" sphere. For the $(N\geq3)$-center metrics, we demonstrated that no symmetry of the Kepler-type occurs for non-aligned centers. We also treated the case of the effective ``truly'' non-Abelian monopole-like field generated by nuclear motion in a diatomic molecule. This system is due to Wilczek et al. where \textit{despite the non-conservation of the electric charge (\ref{noecharge})}, we surprisingly constructed, in addition to the ``unusual'' angular momentum (\ref{diatangmomJ}), a new conserved charge (\ref{Gamma}). 

It is worth mentioning that apart from the generic importance of constructing conserved quantities to confine classical trajectories to conic sections as in the case of Kepler-type systems, the existence of quadratic conserved quantity like Runge-Lenz vector yields, in particular,  the separability of the Hamilton-Jacobi equation for the generalized Taub-NUT and two-center metrics. 
 
``Hidden" symmetries also play a r\^ole in quantum mechanics. The system may be quantized by the usual procedure of replacing Poisson brackets with commutators and in this way the energy levels and degeneracies of Kepler-type systems may be found using dynamical symmetries. 

We mention in conclusion that the van Holten's method can also be extended to study supersymmetries, cf. \cite{Ngome2}.



\begin{acknowledgments}
\noindent
I am indebted to Peter Horv\'athy for valuable discussions on this article. I am also grateful to the SEENET-MTP, Radu Constantinescu, Mihai Visinescu and Sabin Stoica for the invitation to participate to the 8th QFTHS 2012, as well as for the hospitality extended to me in Bucharest and Craiova, Romania. 
\end{acknowledgments}



\end{document}